\documentclass[prl]{revtex4}
\usepackage{graphicx}

\begin{document}

\title{Computer Simulations of Carbon Nanostructures under Pressure}

\author{M. G. Fyta and P. C. Kelires}
\affiliation{Physics Department, University of Crete, P.O. Box 2208, 
710 03, Heraklion, Crete, Greece \\
and Foundation for Research and Technology-Hellas (FORTH), P.O. Box 1527,
711 10 Heraklion, Crete, Greece }


\begin{abstract}
Several interesting phenomena are observed when materials are put
under pressure. The goal is to achieve modification and control over
their mechanical and electronic (conduction) properties. Within this
spirit, we have recently focused our attention into how carbon
nanostructures respond to hydrostatic pressure. We performed
 Monte Carlo simulations with the Tersoff
potential of various free-standing carbon nanostructures. These range
from fullerenes, onions and carbon spheres to nanotubes and nanodiamonds.
Our simulations show that the nanostructures undergo some notable structural 
modifications. 
\end{abstract}
\maketitle

\section{Introduction}

In recent years, novel carbon materials are studied intensively because of 
their enormous potential utility. They display unique structural, mechanical 
and electronic properties. Much research has been done since the mid 80's
when $C_{60}$ was discovered \cite{c60}. New perspectives conjured after
finding the 4th form of Carbon: nanotubes \cite{cnt}. These structures 
 have inspired interesting advances in science and guided the 
visualization and production of similar other stable structures such as
peapods ($C_{60}$ inside CNT) \cite{peapod} and a variety of fullerenes. In these categories of Carbon forms, Carbon onions\cite{onions} and Carbon
 spheres\cite{spheres} are also included.

 Carbon nanotubes (CNT) display a unique structure and exhibit exceptional
 properties such as high stiffness and flexibility, tunable electrical 
conduction between metallic and semiconducting states.
Fullerenes, on the other hand, 
are ideal for manipulating materials on a molecular scale in order to form or
tailor structures of this size. 
Regarding the hybrid forms of CNT and fullerenes, namely peapods, 
their tunable electronic properties must be underlined, since they exhibit
 electronic features additional to that of CNT, which are strongly
 dependent on the location of fullerenes along the tube. In addition, 
Carbon onions can be used in
 energy storage and fuel cells and as immensely tiny ball bearings that
 may be used in nanomachines built on the scale of molecules. Carbon 
spheres \cite{spheres}, which consist of an amorphous Carbon (a-C) 
core and graphine
 (fullerene-like) shells, might find potential application such as
catalyst carriers, lubricants and hydrogen storage materials.

On the other hand, a hybrid $sp^2$ - $sp^3$ Carbon nanocluster 
are bucky diamonds. These are Carbon nanoparticles with a diamond core 
of a few nm and a fullerenelike surface \cite{nanoD}. 
Bucky diamonds are expected to
be present at meteorites in residues of detonation as well as inclusions of
diamondlike films. Quantum confinement effects are present if these bucky
diamonds become smaller than 2 nm. An open, though, question is their
behaviour if they are coalesced, suspended or interconnected.

All of the above nanostructures are potential building blocks of 
nanotechnology. Many of them have been extensively investigated theoretically
and experimentally to discover their properties. Application of pressure
has been used in the case of nanotubes \cite{buckling1}
particularly if they form bundles \cite{buckling2}. 
The pressure, though in most cases is axial, contrary to our calculations.

In the current study, we aim to examine how Carbon nanostructures respond 
under pressure and reveal a transformation path. Hydrostatic pressure is,
thus applied and the structural deformation are examined. Estimations on 
critical values of pressure at which these deformations occur 
will also be presented. An intimation, finally, of the coalescence of 
bucky diamonds under pressure will be made as it is essential to
know the way that nanoscale units join. These issues are interesting as they 
can lead to different configurations and provide ways of modification and control of the electronic properties of the Carbon nanostructures \cite{coal}.

\section{Methodology}

Computer simulations using empirical potentials are effective methods for the
analysis of structural and mechanical properties of complex systems. 
The Monte-Carlo (MC) technique, more specifically, is able to handle large
systems and is preferable for statistical accuracy. A canonical (N,P,T)
ensemble was used in order to apply pressure to the system.
In this approach, the atomic interactions were modeled through
 the Tersoff potential \cite{tersoff}. 

In addition to the Tersoff potential the Lennard-Jones potential was also
implemented into the calculations to simulate the long range Van der Waals
 forces 
wherever they exist (i.e. among $C_{60}$ molecules and nanotubes or a-C atoms 
and nanotube atoms, and more generally between $sp^2$ cages). The parameters
of this potential were given by Lu and Wang \cite{ljparam} :
 $\epsilon$=2.964 meV, $\sigma$=3.407 \AA, and have been used succesfully
 to describe the bulk properties of solid $C_{60}$ and multiwall nanotubes
 \cite{ljparam,ljmwnt})

\section{Results - Discussion}

Hydrostatic pressures ranging from 0 to 100 GPa were applied on Carbon
 nanostructures. As a first step we use only 
one nanostructure per simulation. The simulation temperature is 300 K, 
while calculations for 1500K and 2000K in the case of nanodiamond coalescence
 were also carried out.

The effect of pressure on single wall nanotubes (swnt) has already been studied,
although in most cases the pressure was axial \cite{buckling1,buckling3}. The resulting
structures consist of a series of segments resembling deformed ellipsoids, 
connected with narrowed junctions and are actually buckled. This buckling
is noticeable for pressures higher than 3 GPa and range up to 5-7 Gpa.
Some $sp^3$ sites the approaching shells are also formed. These $sp^3$ 
atoms are mostly clustered and are located on the narrowed junctions.
An elastic recovery of these buckled swnts upon decompression was found,
 although as the pressure increases some defects still exist.

Multi-wall nanotubes (mwnts) under pressure show similar trends, where $sp^3$
hybridization is again visible. These results coincide with another
stuy published while we were analyzing our results\cite{buckling4}. 
The buckling of mwnts is visible at relative high pressures. The critical
pressure is about one order of magnitude higher than in the case of swnts.
It is calculated around  30-50 GPa.
Long range van der Waals (vdW) forces are present between the adjacent
walls of a mwnt. These may play a role in the significant increase of the
 buckling pressure.

An analogous increase of the buckling pressure was evident also in the case of
 compressed peapods. Another point here, is that as pressure continues to
rise the surrounding swnt may bond to the inner array of $C_{60}s$. Focusing
at this inner array shows a coalescence of the fullerene molecules 
while buckling takes place. An example of a compressed (12,12) peapod 
at a pressure of 70 GPa, is shown in Fig.1a. The inner coalesced $C_{60}s$ 
at this pressure is shown in Fig.1b.

Fullerenes, on the other hand, spherical and oval shaped were also put under pressure.
 The former obtain a shape similar to a buckled star, while the latter
a more rectangular one. 
An initially
spherical ($C_{540}$) and an oval shaped ($C_{100}$)  fullerene 
with diameters of about 21 and 11.6 \AA, correspodingly are shown in Fig.2.
Their transformation takes place at modest pressures $\approx$ 5-6 GPa. 

The effect of hydrostatic pressure on Carbon onions is similar to that
of fullerenes and propagates through the shells leaving almost unaffected
the inner shell which in our simulations corresponds to small fullerenes
of approximately 0.8-1nm size, like $C_{60}$ or $C_{70}$. Carbon onions
compressed at around 50 GPa develop a surface similar to that of spherical
onions.

One remark concerning the configuration of Carbon spheres is that the internal 
amorphous Carbon (a-C) forms a shell
resembling that of the fullerene, having a high percentage of 6-fold rings. No 
bonding takes place among the inner and the outer shell. As the pressure 
increases up to approximately 80-100 GPa the spheres, both the fullerene shell 
and the a-C core, flatten at their edges and turn into cubes (Fig.2). 

A final preliminary result that will be given here is the coalescence of
diamond nanocrystals.
 The open question is whether they develop a reconstructed
surface (and become bucky diamonds) even if they are compressed. 
In order to give an answer, we firstly
compressed under various pressures a free standing spherical nanodiamond
of a 6.8 \AA radius. 
 As we raised the temperature up o 1500K the fullerenelike reconstruction
occured even under the effect pressures up to 50-70 GPa. 
The same procedure
was carried out for two diamond nanocrystals at various small distances of  
3-10 \AA.
 The reconstruction again occured at similar to the above, conditions.
An example is given in Fig.4, where two nanodiamonds of a 6.8 \AA radius 
are compressed and their coalescence is promoted. In the case, finally of
relative small temperatures close to 300-500K, the bonding among
the two nanostructures is visible, but no reconstruction occurs.

\section{CONCLUSIONS}

Results regarding the effect of hydrostatic pressure on different
 free-standing Carbon structures of the nanometer scale were presented. Their
structural transformations were studied, which may reveal 
a path to the modification of theis electronic properties.A modest buckling pressure for single 
wall nanotubes were found, which increased significantly for structures
including van der Waals forces. Finally, a preliminary study on the coalescence of nanodiamonds showed that they develop a fullerenelike reconstructed
surface even under pressure.

\newpage

\begin{center}

\begin{figure}
\vspace*{2cm}
\includegraphics[width=0.65\textwidth]{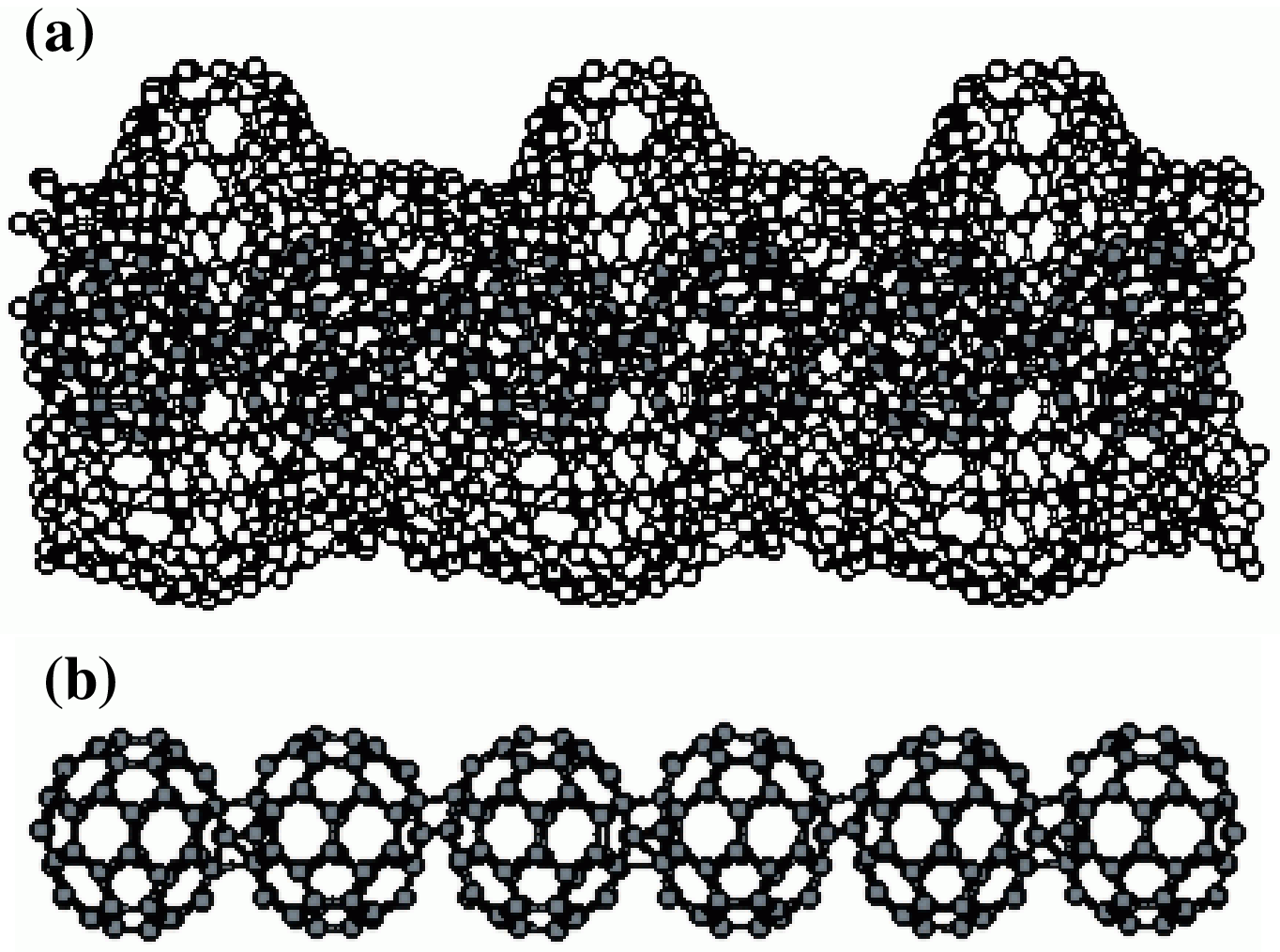}
\caption{Peapod under a pressure of 100GPa (a). The coalesced
array of the internal $C_{60}s$ is shown in (b).}
\end{figure}

\begin{figure}
\vspace*{2cm}
\includegraphics[width=0.5\textwidth]{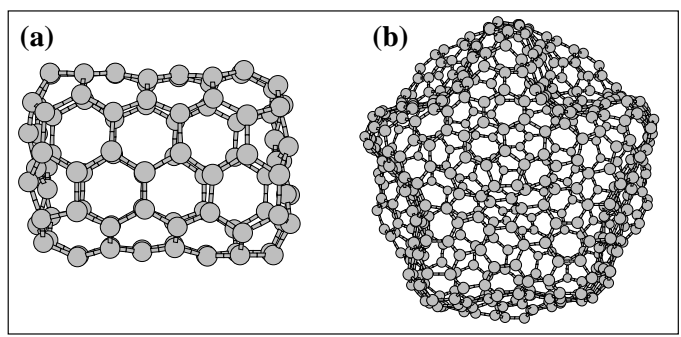}
\caption{Figures of $C_{100}$ and $C_{540}$ after their 
compression at $\approx$ 5 GPa. Their initial shape was oval and spherical,
correspondingly.}
\end{figure}

\begin{figure}
\vspace*{2cm}
\includegraphics[width=0.6\textwidth]{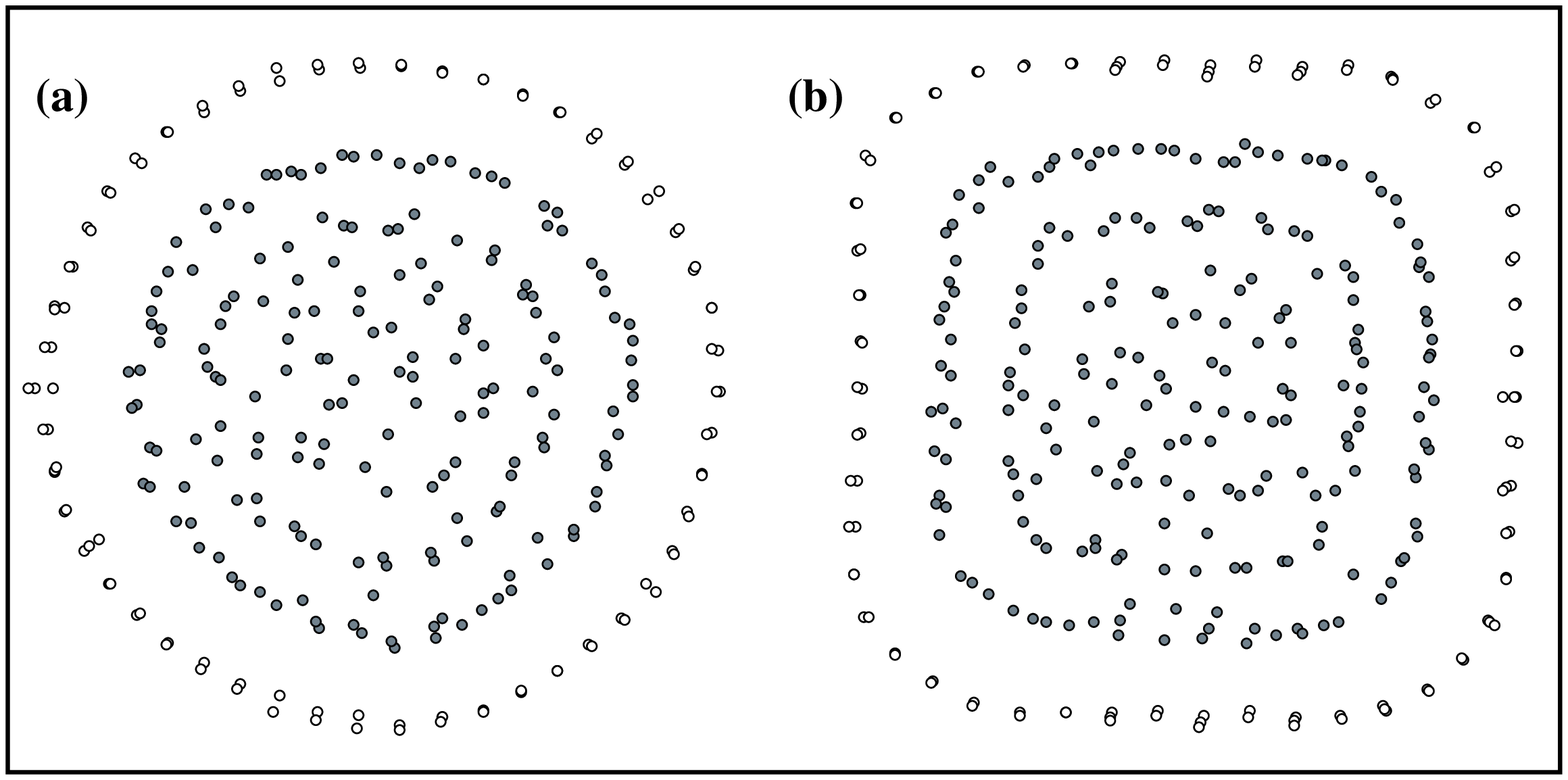}
\caption{Views of Carbon spheres (a) before and (b) after the
exertion of 100 GPa. The internal grey atoms correspond to the a-C core.}
\end{figure}

\begin{figure}
\vspace*{2cm}
\includegraphics[width=0.5\textwidth]{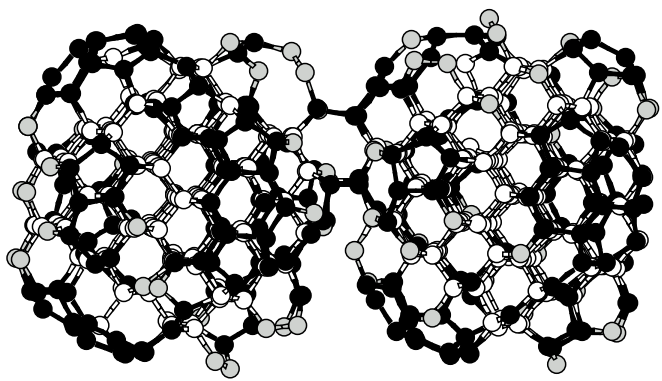}
\caption{Coalescing of 2 nanodiamonds (each with a radius of 6.8 \AA)
at 1500K under a pressure of 20 GPa. White atoms correspond to $sp^3$ bonding,
 while the black ones to $sp^2$. Bonds close to 1.4-1.5 \AA~are formed.}
\end{figure}

\end{center}

\end{document}